# GPUTB-2: An efficient E(3) network method for learning high-precision orthogonal Hamiltonian

Yunlong Wang[1,†], Zhixin Liang[1,†], Chi Ding[1], Junjie Wang[1,*], Zheyong Fan[2], Hui-Tian Wang[1], Dingyu Xing[1], and Jian Sun[1,*]

[1]*National Laboratory of Solid State Microstructures, School of Physics and Collaborative Innovation Center of Advanced Microstructures, Nanjing University, Nanjing, 210093, China*

[2] *College of Physical Science and Technology, Bohai University, Jinzhou, P. R. China.*

# ABSTRACT

Although equivariant neural networks have become a cornerstone for learning electronic Hamiltonians, the intrinsic non-orthogonality of linear combinations of atomic orbitals (LCAO) basis sets poses a fundamental challenge. The computational cost of Hamiltonian orthogonalization scales as $O(N^3)$, which severely hinders electronic structure calculations for large-scale systems containing hundreds of thousands to millions of atoms. To address this issue, we develop GPUTB-2, a framework that learns implicitly orthogonality-preserving Hamiltonians by training directly on electronic band structures. Benefiting from an E(3)-equivariant network accelerated by Gaunt tensor product and SO(2) tensor product layers, GPUTB-2 achieves significantly higher accuracy than GPUTB across multiple benchmark systems. Moreover, GPUTB-2 accurately predicts large-scale electronic structures, including transport properties of temperature-perturbed SnSe and the band structures of magic-angle twisted bilayer graphene. By further integrating this framework with the linear-scaling quantum transport (LSQT) method, we investigate the electronic properties of million-atom amorphous graphene and uncover pressure-induced electronic structure transitions in more complex amorphous silicon. Together, these results establish GPUTB-2 as a high-accuracy and scalable approach for predicting orthogonal Hamiltonians.

[†] These authors contributed equally to this work.
[*]Corresponding author. J.S (jiansun@nju.edu.cn); J.W. (wangjunjie@nju.edu.cn)

# Introduction

Density functional theory (DFT) is one of the most fundamental and well-established theoretical frameworks for electronic structure calculations and materials modeling [1–3]; however, its practical performance for large-scale systems depends critically on the choice of basis set. In conventional plane-wave (PW)–based approaches [4], the Kohn–Sham equations are expanded and iteratively solved using delocalized basis functions defined over the entire simulation cell, leading to a computational complexity that typically scales as $O(N^3)$ with system size. As a result, such methods encounter severe limitations in both computational time and memory consumption when applied to systems containing hundreds or more atoms. In contrast, linear combinations of atomic orbitals (LCAO) [5,6] employ spatially localized atomic orbitals as basis functions, which naturally give rise to sparse Hamiltonian and overlap matrices. This sparsity substantially reduces computational and storage costs and provides a favorable foundation for efficient parallelization as well as near–linear-scaling algorithms [7]. Consequently, LCAO-based methods enable DFT calculations to be extended, with controllable accuracy, to systems comprising thousands of atoms and to long-time ab initio molecular dynamics simulations [8]. Although the completeness of an LCAO basis depends on the choice and number of atomic orbitals—often requiring additional basis-set design and optimization to accurately describe strongly delocalized or high-energy electronic states [9,10]—their pronounced advantages in computational efficiency, resource utilization, and scalability render LCAO schemes highly competitive and widely adopted for large-scale materials modeling [11].

Machine learning atomic-orbital-based Hamiltonians have garnered significant attention in recent years for their ability to dramatically accelerate materials property calculations. Advanced methods such as DeePH [12,13], HamGNN [14] and others have emerged, achieving sub-meV accuracy in Hamiltonian matrix element predictions. However, nearly all existing approaches rely on the introduction of overlap matrices [15,16], necessitating additional orthogonalization costs when solving eigenvalue problems. This limitation results in an overall computational complexity exceeding linear scaling, thereby hindering extension to million-atom systems. Both DeePTB [17] and GPUTB [18] adopt Slater–Koster (SK) parameterizations to learn the SK matrix elements. The introduction of SK parameters guarantees the orthogonality of the resulting Hamiltonian; however, the corresponding network architectures are relatively simple, with a limited number of learnable parameters, which poses challenges for achieving an accurate representation of the Hamiltonian. In GPUTB, for example, a model trained at 300 K that is sufficient to describe III–V semiconductor compounds typically attains a mean absolute error (MAE) of about 10–30 meV. Motivated by these limitations, the goal is to develop a network architecture that is both more accurate and computationally efficient while preserving Hamiltonian orthogonality, enabling straightforward extension to large-scale systems and applicability to more complex materials, such as magic-angle twisted bilayer graphene and large-scale amorphous structures.

Message Passing Neural Networks (MPNNs) update node states through spatially and geometrically aware neighborhood interactions [19,20]. By incorporating E(3) equivariant tensor products, these models maintain precise learning of local atomic information, enabling high-accuracy scalability to large-scale structural computations [21,22]. However, the computational complexity of E(3) equivariant full tensor products scales as $O(L^6)$ [23] (where L denotes the angular momentum order), severely constraining applications of high-order tensor networks. To address this limitation, Passaro & Zitnick sparsified the Clebsch-Gordan coefficients via local coordinate transformations [24], reducing complexity to $O(L^3)$. This approach was successfully integrated into frameworks such as SLEM [25] and DeePH-2 [26]. Concurrently, Luo et al. [27] achieved identical complexity scaling using Gaunt coefficients, differing only by a constant factor. While both methods preserve E(3) equivariance, GTP (Gaunt Tensor Product) bypasses local coordinate transformations, allowing flexible tensor products between multiple equivariant features. Nevertheless, GTP's inability to generate antisymmetric tensors restricts its direct applicability to Hamiltonian predictions.

In this work, we proposes GPUTB-2, an architecture based on an E(3)-equivariant message-passing neural network (MPNN) framework. To achieve an optimal balance between computational efficiency and predictive accuracy, the model employs only a single Gaunt Tensor Product (GTP) layer, followed by an SO(2)-equivariant tensor product layer. The SO(2) layer enables the final Hamiltonian output to simultaneously incorporate symmetric and antisymmetric tensor components. This design principle has been successfully exploited in Hot-

Ham [28], where, with a remarkably small number of trainable parameters, the model achieves state-of-the-art performance among several machine-learning Hamiltonian approaches, including DeepH-E3 [13], DeepH-2 [26], HamGNN [14], and SLEM [25]. Building upon this concept, we further streamline the network architecture to meet the scalability requirements of large-scale systems, resulting in all model with only 0.35 million trainable parameters. Benchmarked against the DeePTB dataset, GPUTB-2 achieves mean absolute errors of a few meV (average 3.3 meV) – significantly surpassing GPUTB (19 meV). Through integration with the Linear-Scaling Quantum Transport (LSQT) method [29,30], we further computed strain-perturbed electronic structures of cubic SiGe with MAE 6 meV. Additionally, our framework validates the disordered silicon electronic transition identified by Volker L. Deringer et al. [31], successfully scaling to million-atom systems while retaining millielectronvolt accuracy. This demonstrates GPUTB-2's capability as a robust new approach for high-accuracy electronic property calculations in large-scale materials systems.

# Result and Discussion

Message Passing Neural Networks (MPNNs) [32–34] excel at capturing complex inter-node relationships and dependencies in a graph by iteratively passing information among nodes; Fig. 1(a) illustrates the network architecture of GPUTB-2, whose initial inputs to the embedding network (Fig. 1(c)) are the element types of nodes i and j ($Z_i$ and $Z_j$), the neighbor distance $|r_{ij}|$, and the spherical harmonics expansion of the neighbor vector $r_{ij}$ ($Y(r_{ij})$). Within the embedding network, these element types are one-hot encoded and passed through a Multi-Layer Perceptron (MLP) layer, and the output is then concatenated with a radial basis function (RBF) expansion of the neighbor distance $|r_{ij}|$; this combined feature subsequently passes through another MLP layer before being fed into an E(3) equivariant layer along with $Y(r_{ij})$ to produce the edge features $e_{ij}$ (between nodes i and j), which are aggregated to form the output node features, thereby successfully encoding the initial structural information. These node features are then passed into a Gaunt tensor product layer (Fig. 1(b)), which uses a scalar tensor product as residual information to enhance the network's performance, and following this, a Layer Normalization (LayerNorm) layer is applied— to maintain E(3) equivariance, this layer only adds a bias parameter to scalars while subtracting the mean from both tensors and scalars; since the Gaunt tensor product layer only considers symmetric tensors, which could violate the equivariance requirement and thus reduce the accuracy, an SO(2) layer is introduced to incorporate antisymmetric tensors, and Fig. 1(d) displays the computational time for the E(3), Gaunt, and SO(2) tensor products at different orders L, where for L > 4, the Gaunt and SO(2) tensor products exhibit faster in time overhead compared to the E(3) tensor product, and the Hamiltonian basically requires L>=4 for d orbitals, so the Gaunt and SO(2) tensor products significantly improving computational efficiency. The learning target of this framework is the band structure, with the overlap matrix implicitly assumed to be the identity matrix, as illustrated in Fig. 1(e). In contrast to other machine-learning Hamiltonian approaches, we refer to this property as implicit orthogonality. It can be proven that, under this assumption, using equivariant neural networks to learn the Hamiltonian matrix is still reasonable. The basis function of the non-orthogonal Hamiltonian is orthogonalized as $\phi'_\mu = \sum_\nu (S^{-1/2})_{\nu\mu} \phi_\nu$, where S is the overlapping matrix, and $\widetilde{H} = S^{-1/2} H S^{-1/2}$ is Löwdin orthogonalization [35] for Hamiltonian. Our method does not directly confirm the form of the basis function and defaults to its orthogonality. We directly construct the E(3) equivariant Hamiltonian and then fit the energy band and optimize the network weight. Under this formulation, the solution of energy eigenvalues does not require any additional treatment of the overlap matrix, making the method naturally compatible with large-scale quantum transport (LSQT) implementations. For more complex systems, such as PbTe with spin–orbit coupling included, the model is able to accurately reproduce the density of states (DOS) obtained from density functional theory (DFT) for perturbed structures, as shown in Fig. 1(f). These results collectively demonstrate the potential of GPUTB-2 for accurate and scalable electronic-structure modeling.

We evaluated several systems listed in Table 1, and the results show that GPUTB-2 consistently achieves significantly higher model accuracy than GPUTB. To further verify that the improved model accuracy translates into enhanced predictive performance for derived physical properties, we trained a temperature-perturbed SnSe model. As shown in Fig. 2(a), the mean absolute error (MAE) of the band structure is as low as 3.2 meV. We then employed PYATB [36] for post-processing of the predicted Hamiltonians to compute the joint density of states (JDOS), shift current, optical conductivity, and dielectric function, and compared these results with those obtained from Hamiltonians derived from density functional theory (DFT) in Fig. 2(b-e). Owing to the high accuracy of the Hamiltonians produced by GPUTB-2, the resulting physical properties show good agreement with the corresponding DFT calculations.

For larger and more complex systems, such as magic-angle twisted bilayer graphene with more than $10^4$ atoms, direct DFT calculations require enormous computational resources. Even when using an LCAO basis, the diagonalization of the Hamiltonian still demands extremely high memory and computational time. We construct our dataset based on relaxed sliding bilayer graphene, as shown in Fig. 3(a). Since the dataset does not explicitly contain twist angles, the larger the system size (corresponding to smaller twist angles), the higher the extrapolation accuracy of the model. In addition to comparing model accuracy, we also benchmark the efficiency of Hamiltonian construction by DFT and GPUTB-2. We generate Hamiltonians for twisted bilayer graphene systems containing up to more than $10^6$ atoms, as shown in Fig. 3(b). At the scale of $10^4$ atoms, the computational time of GPUTB-2 is significantly lower than that of the DFT approach. As shown in Fig. 3(c), we employ the shift-and-invert Krylov subspace iterative algorithm [37] to compute a subset of energy bands near the Fermi level. The computational complexity of the shift-and-invert Krylov method is dominated by the linear system solution in each iteration. As analyzed in Saad [38] and Ericsson & Ruhe [39], the spectral transformation accelerates convergence significantly, reducing the number of iterations. However, the cost per iteration is determined by the factorization of the sparse Hamiltonian. According to Davis [40], for a discretized 3D system, this factorization scales as $O(N^2)$, which is superior to the $O(N^3)$ of full diagonalization but can still be prohibitive for extremely large N. This algorithm efficiently obtains selected eigenvalues with relatively low computational complexity and memory requirements. We further compare the memory consumption and computational time for eigenvalue solutions using the non-orthogonal Hamiltonians generated by DFT (including both the Hamiltonian and overlap matrices) and the orthogonal Hamiltonians generated by GPUTB-2 (containing only the Hamiltonian, with the overlap matrix being the identity). The memory usage and computational time for solving the eigenvalue problem with GPUTB-2 are both significantly lower than those of the DFT method owing to its sparser Hamiltonian matrixes.

To investigate the performance of GPUTB-2 in amorphous systems, we first trained a 300K temperature perturbation model of graphene. Based on this optimized model, we then incorporated a training set containing Stone–Wales (SW) defects to enhance the model's ability to recognize such defects. As illustrated in Fig. 4(a), in the electronic density of states (DOS), the electronic valleys near the Fermi level become filled in the amorphous, disordered structure, producing a smoother overall energy distribution. The structural disorder in the amorphous state is reflected in the radial distribution function, as shown in Fig. 4(b), where the isolated peaks at long range evolve into a continuous profile and gradually approach 1. We also computed the time-dependent conductivity, as shown in Fig. 4(c), averaging over a 10 fs window after convergence at 100 fs. Owing to enhanced electron scattering caused by structural disorder, amorphous graphene exhibits a pronounced suppression of conductivity across the entire energy range. To further demonstrate LSQT's capability for computing electronic properties in large-scale systems, we calculated the density of states (DOS) of amorphous silicon under high pressure. Deringer et al. employed Kwon's tight-binding Hamiltonian (based on *s* and *p* orbitals) and a linear-scaling maximum entropy method to efficiently extract electronic DOS from the sparse Hamiltonian matrix of a $10^5$ atoms system. They also developed a machine learning regression model that rapidly predicts DOS for large systems using only atomic coordinates as input, with results consistent with experimentally observed conductivity jumps. As shown in Fig. 4(d). Using 64 atoms configurations of *cd*, *β*, *sh*, and *Imma* phases Si for training—where the *cd* phase exhibited lower energy than others—our training set extended to 1600 K (near Si's melting point) to encompass diverse configurations and complex atomic environments, achieving a model MAE of 7.4 meV. LSQT calculations on Deringer's configurations yielded consistent results. Notably, while LSQT computes DOS by solving the Hamiltonian, Deringer's method learns DOS directly, potentially omitting implicit physical information. To address this, they used eight committee models to represent methodological uncertainty, reporting averaged DOS—a system introducing significant predictive uncertainty. We trained 8 models with different initialization parameters. When the models converged to the same accuracy, we calculated the uncertainty range of the network's predicted DOS. LSQT's DOS uncertainty decreases sharply with system size as $\Delta\rho \propto \frac{1}{\sqrt{NN_r}}$ [29], where $N$ is the orbital count and $N_r$ is the number of stochastic observations. For $NN_r = 10^4$, $\Delta\rho \approx 1\%$; in our system $NN_r$ is $3.6 \times 10^{10}$, $\Delta\rho \sim 1.67 \times 10^{-4}\%$, rendering LSQT errors negligible. In Fig. 4(e), we annotate the uncertainty of the GPUTB-2 framework. Our results fall within Deringer's uncertainty bounds but the confidence intervals are more accurate. For multiple phases at 12, 13, and 20 GPa, GPUTB-2 consistently produced results analogous to Deringer et al.(Reference Supplementary Materials).

# Discussion and conclusion

In summary, we present GPUTB-2, an E(3)-equivariant message-passing framework designed to learn implicitly orthogonal tight-binding Hamiltonians directly from electronic band structures, thereby eliminating the need for explicit Hamiltonian orthogonalization and its prohibitive $O(N^3)$ computational cost. The proposed methodology integrates a compact yet expressive network architecture that combines a single Gaunt tensor-product layer with an SO(2)-equivariant tensor product, enabling the accurate reconstruction of both symmetric and antisymmetric tensor components while maintaining high computational efficiency. By employing environment-aware embeddings and radial basis expansions, the model effectively captures local chemical and structural information, and the use of band energies as training targets naturally enforces orthogonality, yielding Hamiltonians that are immediately compatible with linear-scaling quantum transport (LSQT) algorithms. Comprehensive benchmarks across a range of crystalline and disordered systems demonstrate that GPUTB-2 consistently achieves state-of-the-art accuracy, with band-energy mean absolute errors on the order of a few meV, significantly outperforming its predecessor. Beyond band structures, the model reliably reproduces derived electronic properties under temperature perturbations, including density of states, joint density of states, optical conductivity, dielectric response, and nonlinear shift currents. Moreover, when combined with LSQT, GPUTB-2 enables efficient simulations of systems containing up to millions of atoms, as illustrated by pressure-induced electronic transitions in amorphous silicon and large-scale amorphous materials. Overall, this work establishes GPUTB-2 as a robust and scalable approach for constructing orthogonal Hamiltonians with high fidelity, offering a practical pathway toward device-scale electronic-structure and transport simulations and opening new opportunities for large-scale, data-driven materials modeling.

# Methods

## Perturbation structure calculation method

The perturbed structure was simulated using molecular dynamics (MD) through GPUMD package [41–44] under the NPT ensemble with a Berendsen thermostat [45]. The simulation employed a time step of 0.5 fs and was conducted for 100 ps. The NEP [46] force field used for the MD calculations was trained based on the results of first-principles calculations performed with ABACUS [47–49]. The training configurations were obtained from NEP89 [50] force field within the same temperature range.

## Training set calculation details

The training set was computed using ABACUS [47–49] with a double-zeta plus polarization (DZP) basis set in the linear combination of atomic orbitals (LCAO) framework and the SG15 ONCV pseudopotentials [51]. The Perdew-Burke-Ernzerhof (PBE) functional, with an energy cutoff of 100 Ry, a k-point spacing of 0.1 1/bohr, and a charge density convergence threshold of $10^{-8}$ Ry. Band structure calculations were performed along standard high-symmetry k-paths. For amorphous silicon, a spatial grid with constant spacing of 0.1 in each direction was utilized.

## Spin orbit coupling effect

In systems with considering spin-orbit coupling (SOC) effects, the system's Hamiltonian incorporating SOC ($H_{soc}$) can be expressed as:

$$H_{soc} = I_2 \otimes H_{no-soc} + \sum_i \lambda_i L_i \cdot S_i \qquad (1)$$

Here, $I_2$ denotes the 2×2 identity matrix, $H_{no-soc}$ represents the Hamiltonian without SOC, $\lambda_i$ signifies the SOC strength parameter, and $L_i$ and $S_i$ are the orbital and spin angular momentum operators for the *i*-th orbital, respectively. Node features are input into the soc fully connected E(3) network and then output $\lambda_i$. The summation extends over all relevant orbitals *i*. Importantly, models utilizing this form generally achieve high fidelity in reproducing high-precision electronic band structures when SOC is included.

# References

[1] L. E. Ratcliff, S. Mohr, G. Huhs, T. Deutsch, M. Masella, and L. Genovese, Challenges in large scale quantum mechanical calculations, WIREs Comput. Mol. Sci. **7**, e1290 (2017).

[2] G. Galli and M. Parrinello, Large scale electronic structure calculations, Phys. Rev. Lett. **69**, 3547 (1992).

[3] M. G. Medvedev, I. S. Bushmarinov, J. Sun, J. P. Perdew, and K. A. Lyssenko, Density functional theory is straying from the path toward the exact functional, Science **355**, 49 (2017).

[4] G. Kresse and J. Furthmüller, Efficiency of ab-initio total energy calculations for metals and semiconductors using a plane-wave basis set, Comput. Mater. Sci. **6**, 15 (1996).

[5] T. Ozaki and H. Kino, Efficient projector expansion for the *ab initio* LCAO method, Phys. Rev. B **72**, 045121 (2005).

[6] D. Sánchez-Portal, P. Ordejón, E. Artacho, and J. M. Soler, Density-functional method for very large systems with LCAO basis sets, Int. J. Quantum Chem. **65**, 453 (1997).

[7] J.-L. Fattebert and J. Bernholc, Towards grid-based O (N) density-functional theory methods: Optimized nonorthogonal orbitals and multigrid acceleration, Phys. Rev. B **62**, 1713 (2000).

[8] G. Seifert, D. Porezag, and T. Frauenheim, Calculations of molecules, clusters, and solids with a simplified LCAO-DFT-LDA scheme, Int. J. Quantum Chem. **58**, 185 (1996).

[9] R. Evarestov, E. Kotomin, Y. A. Mastrikov, D. Gryaznov, E. Heifets, and J. Maier, Comparative density-functional LCAO and plane-wave calculations of La Mn O 3 surfaces, Phys. Rev. B—Condensed Matter Mater. Phys. **72**, 214411 (2005).

[10] T. Ozaki and H. Kino, Efficient projector expansion for the ab initio LCAO method, Phys. Rev. B—Condensed Matter Mater. Phys. **72**, 045121 (2005).

[11] J. C. Slater and G. F. Koster, Simplified LCAO Method for the Periodic Potential Problem, Phys. Rev. **94**, 1498 (1954).

[12] H. Li, Z. Wang, N. Zou, M. Ye, R. Xu, X. Gong, W. Duan, and Y. Xu, Deep-learning density functional theory Hamiltonian for efficient ab initio electronic-structure calculation, Nat.

Comput. Sci. **2**, 367 (2022).

[13] X. Gong, H. Li, N. Zou, R. Xu, W. Duan, and Y. Xu, General framework for E(3)-equivariant neural network representation of density functional theory Hamiltonian, Nat. Commun. **14**, 2848 (2023).

[14] Y. Zhong, H. Yu, M. Su, X. Gong, and H. Xiang, Transferable equivariant graph neural networks for the Hamiltonians of molecules and solids, Npj Comput. Mater. **9**, 182 (2023).

[15] M. J. Han, T. Ozaki, and J. Yu, O (N) LDA+ U electronic structure calculation method based on the nonorthogonal pseudoatomic orbital basis, Phys. Rev. B—Condensed Matter Mater. Phys. **73**, 045110 (2006).

[16] D. Porezag, T. Frauenheim, T. Köhler, G. Seifert, and R. Kaschner, Construction of tight-binding-like potentials on the basis of density-functional theory: Application to carbon, Phys. Rev. B **51**, 12947 (1995).

[17] Q. Gu, Z. Zhouyin, S. K. Pandey, P. Zhang, L. Zhang, and W. E, Deep learning tight-binding approach for large-scale electronic simulations at finite temperatures with ab initio accuracy, Nat. Commun. **15**, 6772 (2024).

[18] Y. Wang, Z. Liang, C. Ding, J. Wang, Z. Fan, H.-T. Wang, D. Xing, and J. Sun, GPUTB: Efficient machine learning tight-binding method for large-scale electronic properties calculations, Comput. Mater. Today **8**, 100039 (2025).

[19] M. Gori, G. Monfardini, and F. Scarselli, *A New Model for Learning in Graph Domains*, in *Proceedings. 2005 IEEE International Joint Conference on Neural Networks, 2005.*, Vol. 2 (IEEE, Montreal, Que., Canada, 2005), pp. 729–734.

[20] J. Gilmer, S. S. Schoenholz, P. F. Riley, O. Vinyals, and G. E. Dahl, *Neural Message Passing for Quantum Chemistry*, in *Proceedings of the 34th International Conference on Machine Learning*, edited by D. Precup and Y. W. Teh, Vol. 70 (PMLR, 2017), pp. 1263–1272.

[21] T. Cohen and M. Welling, *Group Equivariant Convolutional Networks*, in *International Conference on Machine Learning* (PMLR, 2016), pp. 2990–2999.

[22] R. Kondor and S. Trivedi, *On the Generalization of Equivariance and Convolution in Neural Networks to the Action of Compact Groups*, in *International Conference on Machine Learning* (PMLR, 2018), pp. 2747–2755.

[23] Y. Li, X. Zhang, and L. Shen, A critical review of machine learning interatomic potentials and Hamiltonian, J. Mater. Inform. **5**, N (2025).

[24] S. Passaro and C. L. Zitnick, *Reducing SO(3) Convolutions to SO(2) for Efficient Equivariant GNNs*, in *Proceedings of the 40th International Conference on Machine Learning*, edited by A. Krause, E. Brunskill, K. Cho, B. Engelhardt, S. Sabato, and J. Scarlett, Vol. 202 (PMLR, 2023), pp. 27420–27438.

[25] Z. Zhouyin, Z. Gan, S. K. Pandey, L. Zhang, and Q. Gu, *Learning Local Equivariant Representations for Quantum Operators*, in *The Thirteenth International Conference on Learning Representations* (2025).

[26] Y. Wang, H. Li, Z. Tang, H. Tao, Y. Wang, Z. Yuan, Z. Chen, W. Duan, and Y. Xu, *DeepH-2: Enhancing Deep-Learning Electronic Structure via an Equivariant Local-Coordinate Transformer*, arXiv:2401.17015.

[27] S. Luo, T. Chen, and A. S. Krishnapriyan, *Enabling Efficient Equivariant Operations in the Fourier Basis via Gaunt Tensor Products*.

[28] Z. Liang, Y. Wang, C. Ding, J. Wang, H.-T. Wang, D. Xing, and J. Sun, Hot-Ham: an Accurate and Efficient E(3)-Equivariant Machine-Learning Electronic Structures Calculation Framework, Chin. Phys. Lett. **43**, 020704 (2026).

[29] Z. Fan, J. H. Garcia, A. W. Cummings, J. E. Barrios-Vargas, M. Panhans, A. Harju, F. Ortmann, and S. Roche, Linear scaling quantum transport methodologies, Phys. Rep. **903**, 1 (2021).

[30] Z. Fan, Y. Xiao, Y. Wang, P. Ying, S. Chen, and H. Dong, Combining linear-scaling quantum transport and machine-learning molecular dynamics to study thermal and electronic transports in complex materials, J. Phys. Condens. Matter **36**, 245901 (2024).

[31] V. L. Deringer, N. Bernstein, G. Csányi, C. Ben Mahmoud, M. Ceriotti, M. Wilson, D. A. Drabold, and S. R. Elliott, Origins of structural and electronic transitions in disordered silicon, Nature **589**, 7840 (2021).

[32] J. Gilmer, S. S. Schoenholz, P. F. Riley, O. Vinyals, and G. E. Dahl, *Neural Message Passing for Quantum Chemistry*, in *International Conference on Machine Learning* (Pmlr, 2017), pp. 1263–1272.

[33] I. Batatia, D. P. Kovacs, G. Simm, C. Ortner, and G. Csányi, MACE: Higher order equivariant message passing neural networks for fast and accurate force fields, Adv. Neural Inf. Process. Syst. **35**, 11423 (2022).

[34] J. Gilmer, S. S. Schoenholz, P. F. Riley, O. Vinyals, and G. E. Dahl, *Message Passing Neural Networks*, in *Machine Learning Meets Quantum Physics* (Springer, 2020), pp. 199–214.

[35] P.-O. Löwdin, *On the Nonorthogonality Problem*, in *Advances in Quantum Chemistry*, Vol. 5 (Elsevier, 1970), pp. 185–199.

[36] G. Jin, H. Pang, Y. Ji, Z. Dai, and L. He, PYATB: An efficient Python package for electronic structure calculations using ab initio tight-binding model, Comput. Phys. Commun. **291**, 108844 (2023).

[37] R. B. Lehoucq, D. C. Sorensen, and C. Yang, *ARPACK Users' Guide: Solution of Large-Scale Eigenvalue Problems with Implicitly Restarted Arnoldi Methods* (SIAM, 1998).

[38] Y. Saad, *Numerical Methods for Large Eigenvalue Problems: Revised Edition* (SIAM, 2011).

[39] T. Ericsson and A. Ruhe, The spectral transformation Lanczos method for the numerical solution of large sparse generalized symmetric eigenvalue problems, Math. Comput. **35**, 1251 (1980).

[40] T. A. Davis, *Direct Methods for Sparse Linear Systems* (SIAM, 2006).

[41] Z. Fan et al., GPUMD: A package for constructing accurate machine-learned potentials and performing highly efficient atomistic simulations, J. Chem. Phys. **157**, 114801 (2022).

[42] Z. Fan, Z. Zeng, C. Zhang, Y. Wang, K. Song, H. Dong, Y. Chen, and T. Ala-Nissila, Neuroevolution machine learning potentials: Combining high accuracy and low cost in atomistic simulations and application to heat transport, Phys. Rev. B **104**, 104309 (2021).

[43] K. Xu et al., GPUMD 4.0: A high-performance molecular dynamics package for versatile materials simulations with machine-learned potentials, Mater. Genome Eng. Adv. e70028 (2025).

[44] Z. Fan, W. Chen, V. Vierimaa, and A. Harju, Efficient molecular dynamics simulations with many-body potentials on graphics processing units, Comput. Phys. Commun. **218**, 10

(2017).

[45] H. J. Berendsen, J. van Postma, W. F. Van Gunsteren, A. DiNola, and J. R. Haak, Molecular dynamics with coupling to an external bath, J. Chem. Phys. **81**, 3684 (1984).

[46] K. Song et al., General-purpose machine-learned potential for 16 elemental metals and their alloys, Nat. Commun. **15**, 10208 (2024).

[47] P. Lin, X. Ren, X. Liu, and L. He, Ab initio electronic structure calculations based on numerical atomic orbitals: Basic fomalisms and recent progresses, WIREs Comput. Mol. Sci. **14**, e1687 (2024).

[48] P. Li, X. Liu, M. Chen, P. Lin, X. Ren, L. Lin, C. Yang, and L. He, Large-scale ab initio simulations based on systematically improvable atomic basis, Comput. Mater. Sci. **112**, 503 (2016).

[49] P. Lin, X. Ren, and L. He, Strategy for constructing compact numerical atomic orbital basis sets by incorporating the gradients of reference wavefunctions, Phys. Rev. B **103**, 235131 (2021).

[50] T. Liang et al., NEP89: Universal neuroevolution potential for inorganic and organic materials across 89 elements, ArXiv Prepr. ArXiv250421286 (2025).

[51] M. Schlipf and F. Gygi, Optimization algorithm for the generation of ONCV pseudopotentials, Comput. Phys. Commun. **196**, 36 (2015).

# Acknowledgements

We gratefully acknowledges the National Natural Science Foundation of China (Grant 12125404, T2495231, 123B2049), the Basic Research Program of Jiangsu (Grant BK20233001, BK20241253), the Jiangsu Funding Program for Excellent Postdoctoral Talent (Grants 2024ZB002 and 2024ZB075), the Postdoctoral Fellowship Program of CPSF (Grant GZC20240695), the AI & AI for Science program of Nanjing University, Artificial Intelligence and Quantum physics (AIQ) program of Nanjing University, and the Fundamental Research Funds for the Central Universities. The calculations were carried out using supercomputers at the High-Performance Computing Center of Collaborative Innovation Center of Advanced Microstructures, the high-performance supercomputing center of Nanjing University.

# Figures

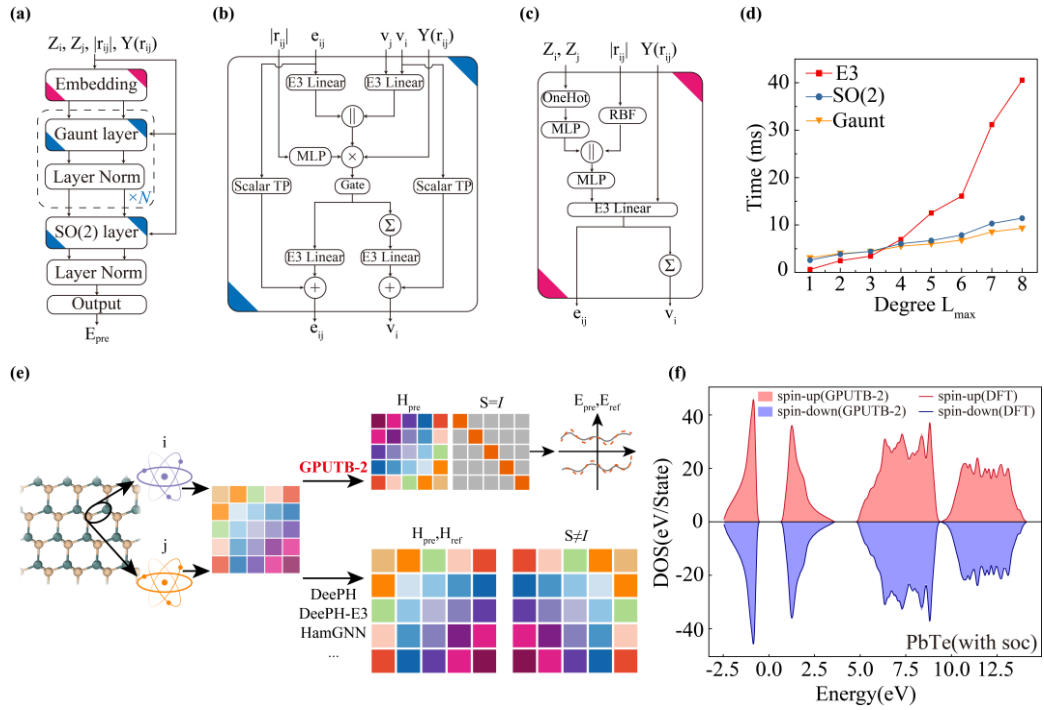

**Fig. 1. Flowchart and network architecture of GPUTB-2. (a)** The main architecture and training flow chart of GPUTB-2. **(b)** The tensor product network for Gaunt and SO(2) layer. **(c)** The embedding network, where RBF is radial basis function, $e_{ij}$ and $v_i$ are the edge and node feature respectively. **(d)** Comparison of calculation time of E3, Gaunt and SO(2) tensor product under different expansion orders. **(e)** Different from methods such as DeePH, DeePH-E3, and HamGNN that directly learn Hamiltonians, GPUTB-2 learns energy bands to ensure the orthogonality of Hamiltonians. **(f)** Spin-up and spin-down electronic state density of PbTe considering spin-orbit coupling.

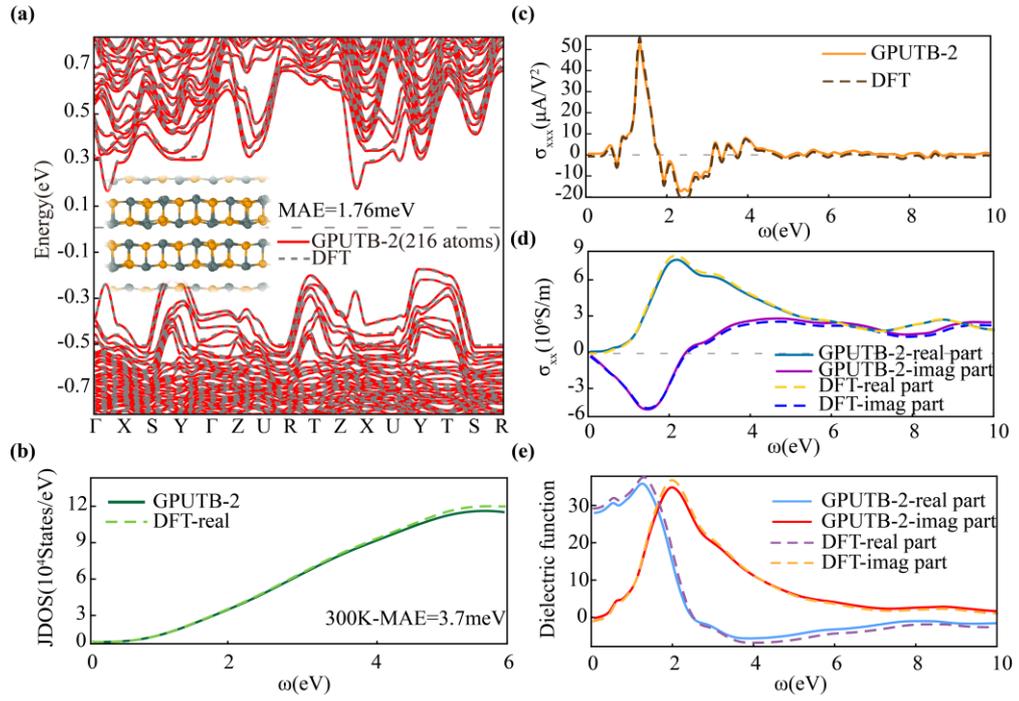

**Fig. 2. Predictions of Hamiltonian properties of 256 atoms SnSe under temperature perturbation.** **(a)** Generalization testing, training 64 atom SiGe under temperature perturbation, predicting the accuracy of 216 atom SiGe. **(b)** Joint density of states (JDOS) compared to DFT data. **(c)** The shift current compared to DFT data. **(d)** The optical conductivity compared with DFT data. **(e)** The dielectric function compared with DFT data.

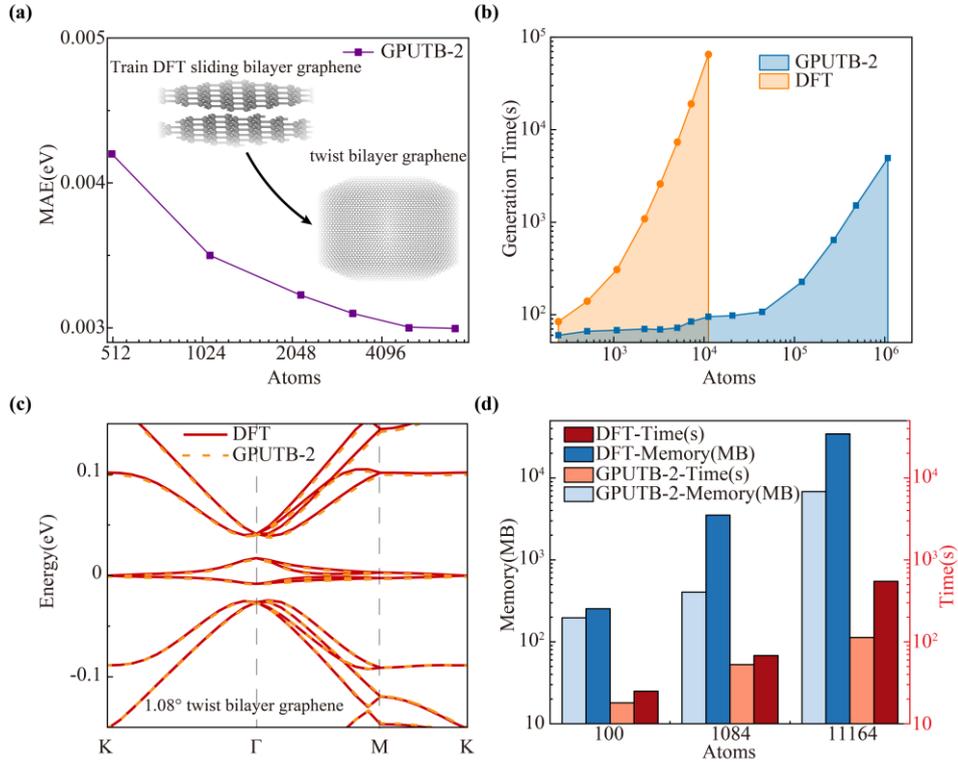

**Fig. 3. Predicting the energy bands of magic angle graphene near the Fermi surface. (a)** Trend of the mean absolute error (MAE) of the band energies within ±0.5 eV relative to the Fermi level, evaluated on a dataset of untwisted slid bilayer graphene. **(b)** Performance comparison of Hamiltonian generation in GPUTB-2 and DFT for the same structure. **(c)** Comparison of the electronic band structure near the Fermi level of magic-angle graphene calculated using GPUTB-2 and DFT. **(d)** Comparison of computational time and memory overhead between orthogonal and non-orthogonal Hamiltonians in solving generalized eigenvalue problems using Krylov subspace iterative algorithms.

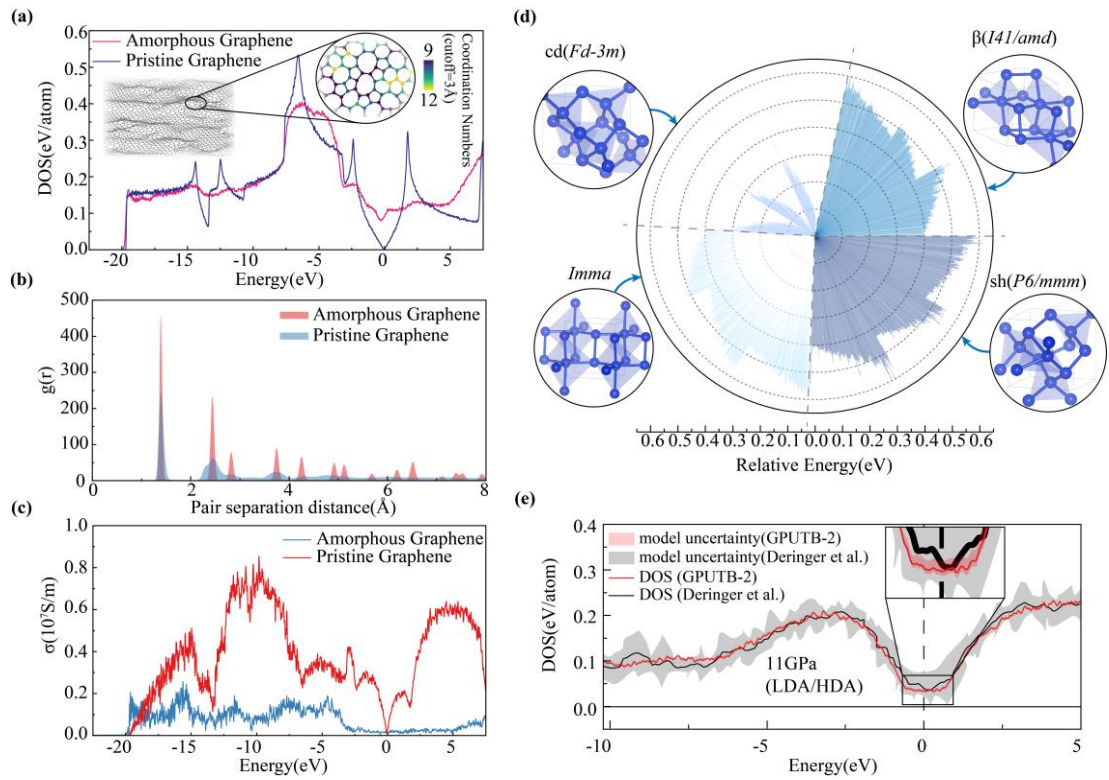

**Fig. 4. GPUTB-2 accurately reproduces the electronic structure of amorphous graphene and high-pressure amorphous silicon.** **(a)** The DOS of amorphous graphene and pristine graphene. **(b)** The radial distribution function of amorphous graphene and pristine graphene. **(c)** The conductivity of amorphous graphene and pristine graphene. **(d)** The structure of the training set and its energy distribution, the relative energy is all configurations minus the minimum energy of the training set. **(e)** The DOS of amorphous silicon under 11GPa are compared with Deringer et al.

# Data availability

Data will be made available on request.

# Tables

Table. 1. Comparison the results of GPUTB and GPUTB-2.

|  | Diamond | Silicon | SiGe | GaP | | | AlAs | | | Average |
|---|---|---|---|---|---|---|---|---|---|---|
|  | c/c | c/c | c/c | c/c | m/c | m/h | c/c | m/c | m/h |  |
| GPUTB | 16 | - | 13 | 12 | 12 | 22 | 20 | 21 | 35 | 19 |
| **GPUTB-2** | **2.0** | **1.4** | **3.5** | **2.2** | **3.2** | **7.2** | **2.3** | **2.9** | **5.2** | **3.3** |

MAEs are in unit of meV, c/c represents using cubic phase training and testing, m/c represents using mixed phase training and cubic phase testing, m/h represents using mixed phase training and hexagonal phase testing.

Table. 2. Dataset details and MAE of the other systems discussed in the manuscript.

| System | Space Group | Dataset structures | Dataset atoms | Temperature | MAE(meV) |
|---|---|---|---|---|---|
| SnSe | Pnma | 975 | 64 | 300K | 3.20 |
| Bilayer graphene | P6/mmm | 491 | 100 | 0K | 0.8 |
| Amorphous silicon | Fd-3m, I41/amd, Imma, P6/mmm | 1800 | 64 | 400-1600K | 7.46 |
| Amorphous graphene | Random Stone-Wales defect | 1272 | 72 | 300K | 4.75 |